\documentclass[12pt]{article}
\usepackage[colorlinks=true, urlcolor=navyblue, linkcolor=navyblue, citecolor=navyblue]{hyperref}
\usepackage{graphicx,amsmath,amssymb}
\usepackage{indentfirst}
\usepackage[usenames]{color}
\usepackage{enumerate}
\usepackage{appendix}
\usepackage{setspace}

\definecolor{navyblue}{rgb}{0,0.08,0.45}
\definecolor{darkred}{rgb}{0.7,0.0,0.0}
\definecolor{darkgreen}{rgb}{0,0.6,0.2}

\begin{document}


\begin{flushright}
{\small SLAC--PUB--16397 \\ \vspace{2pt}}
\end{flushright}

\vspace{30pt}

\begin{center}
{\huge Light-Front Holography and Superconformal}

\end{center}

\vspace{-20pt}

\begin{center}
{\huge Quantum Mechanics: A New Approach to}
\end{center}

\vspace{-20pt}

\begin{center}
{\huge Hadron Structure and Color Confinement}
\end{center}

\centerline{Stanley J. Brodsky}

\vspace{3pt}

\centerline {\it SLAC National Accelerator Laboratory, Stanford
University, Stanford, CA 94309,
USA\,\footnote{\href{mailto:sjbth@slac.stanford.edu}{\tt
sjbth@slac.stanford.edu}}}

\vspace{8pt}

\centerline{Alexandre Deur}

\vspace{3pt}

\centerline{\it Thomas Jefferson National Accelerator Facility, Newport
News, VA 23606, USA\,\footnote{{\href{mailto:deurpam@jlab.org}{\tt
deurpam@jlab.org}}}}

\vspace{8pt}

\centerline{Guy F. de T\'eramond}

\vspace{3pt}

\centerline {\it Universidad de Costa Rica, 11501 San Jos\'e, Costa
Rica\,\footnote{{\href{mailto:gdt@asterix.crnet.cr}{\tt
gdt@asterix.crnet.cr}}}}

\vspace{8pt}

\centerline{Hans G\"unter Dosch}

\vspace{3pt}

\centerline {\it Institut f\"ur Theoretische Physik, Philosophenweg
16, D-6900 Heidelberg\,\footnote{{\href{mailto:h.g.dosch@thphys.uni-heidelberg.de}{\tt
h.g.dosch@thphys.uni-heidelberg.de}}}}

\vspace{40pt}

\date{\today}

\begin{abstract}
A primary question in hadron physics  is how the mass scale for hadrons consisting of light quarks, such as the proton,  emerges from the QCD Lagrangian even in the limit of zero quark mass. 
If one requires the  
effective action which underlies the  QCD Lagrangian to remain conformally invariant and extends the formalism of de Alfaro, Fubini and Furlan to light-front Hamiltonian theory, then a unique,  color-confining  potential with a mass parameter $\kappa$ emerges. The actual value of the parameter $\kappa$ is not set by the model -- only ratios of hadron masses and other hadronic mass scales are predicted. 
The result is a nonperturbative, relativistic light-front quantum mechanical wave equation, the {\it Light-Front Schr\"odinger Equation } which incorporates color confinement and other essential spectroscopic and dynamical features of hadron physics, including a massless pion for zero quark mass and linear Regge trajectories 
with the identical slope in the 
radial quantum number $n$ and orbital angular momentum $L$.   
The same light-front  equations for mesons with spin $J$ also can be derived from the holographic mapping to QCD (3+1) at fixed light-front time
from the soft-wall model modification of AdS$_5$ space with a specific dilaton profile.
Light-front holography thus provides a precise relation between the bound-state amplitudes in the fifth dimension of AdS space and the boost-invariant light-front wavefunctions describing the internal structure of hadrons in physical space-time.   
One can also extend the analysis to baryons using superconformal algebra -- $2 \times 2 $ supersymmetric  representations of the conformal group.  The resulting fermionic  LF bound-state equations predict striking
similarities between the meson and baryon spectra.  In fact,
the holographic QCD light-front Hamiltonians
for the states on the meson and baryon trajectories are identical
if one shifts the internal  angular momenta of the meson ($L_M$)
and baryon ($L_B$)  by one unit: $L_M=L_B+1$.  
We  also show how the mass scale $\kappa$ underlying confinement and the masses  of light-quark hadrons determines the scale  
 $\Lambda_{\overline{MS}}$  controlling the evolution of the perturbative QCD coupling.  The relation between scales is obtained by matching the nonperturbative dynamics, as described by an effective conformal theory mapped to the light-front and its embedding in AdS space, to the perturbative QCD  regime. The data for the effective coupling defined from the Bjorken sum rule $\alpha_{g_1}(Q^2)$  are remarkably consistent with the  Gaussian form  predicted by LF holographic QCD.
The result is an effective coupling  defined at all momenta. The predicted value $\Lambda^{(N_F=3)}_{\overline{MS}} = 0.440 m_\rho = 0.341 \pm 0.024$ GeV is in agreement with the world average $0.339 \pm 0.010$ GeV.  We thus can connect $\Lambda_{\overline{MS}}$ to hadron masses.  The analysis applies to any renormalization scheme.

\end{abstract}

\newpage


\section{Emergence of the QCD Mass Scale}
A central feature of QCD for zero quark mass is the appearance of the hadronic mass scale, even though no parameter with mass dimension appears explicitly in the QCD Lagrangian. In fact, at the semiclassical level, QCD  with massless quarks is conformally invariant.    

The emergence of a mass scale in an underlying conformal theory was investigated in the context of one-dimensional quantum mechanics in 1976 in a pioneering paper by de Alfaro, Fubini,  and Furlan (dAFF)~\cite{deAlfaro:1976je}.
They showed that one can modify the Hamiltonian of a conformal theory with a new term containing a constant $\kappa$ with the dimension of mass while retaining the conformal invariance of the action 
by including contributions proportional to the dilatation and special conformal operators of the conformal group.   
The result is a new term in the Hamiltonian which has the unique form of a  confining harmonic oscillator potential.  
Thus not only does a mass  scale emerge, but the spectrum only consists of bound states.  Remarkably, the action remains conformally invariant if one transforms the time variable from ordinary time $t$  to a  new evolution parameter $\tau$ which can be interpreted as the light-front time $t+z/c$, the time along a light-front (LF).
The range of $\tau$ is finite reflecting the finite difference of light-front times between the confined constituents.~\cite{Brodsky:2013ar}

\begin{figure}[h]
\centering
\includegraphics[width=.90\textwidth]{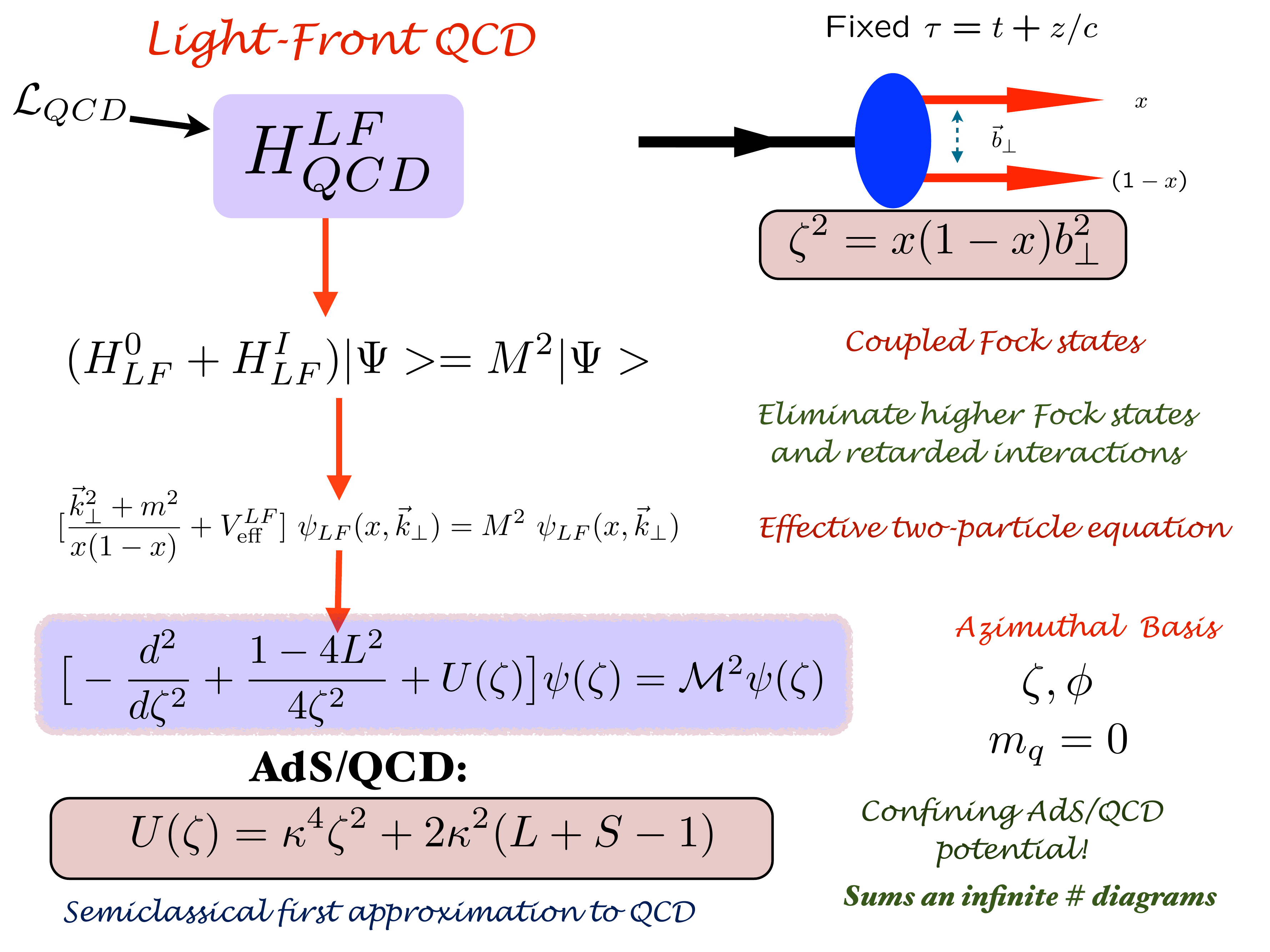} \vspace{10pt}
\caption{\label{SlovakiaReduction}  
Systematic reduction of the full LF Hamiltonian of QCD to an  effective LF Schr\"odinger equation for the $q \bar q$ valence meson Fock state.  }
\end{figure}

The dAFF procedure can be applied to hadron physics~\cite{Brodsky:2013ar,Dosch:2014wxa} in physical (3+1) space-time in the context of  light-front quantization; i.e., Dirac's ``Front Form".  
One can define the LF Hamiltonian 
$H_{LF} = - i {d\over d\tau}$ directly from the QCD Lagrangian. The eigensolutions of the LF Hamlitonian predict the hadronic mass spectrum and the frame-independent light-front  wavefunctions which underly hadronic observables -- form factors, structure functions, transverse momentum distributions, etc. 
As shown in Fig. \ref{SlovakiaReduction}, the exact LF Heisenberg equation $H_{LF} \Psi = M^2|\Psi>$ for mesons can be systematically reduced to an effective {\it LF Schr\"odinger Equation } (LFSE)  operating 
on the lowest $q \bar q$ Fock state. (In the case of QED, similar steps generate the effective  bound state equation  for hydrogenic atoms, including the Lamb Shift and hyperfine splitting.) The  resulting  LF  equation  is frame-independent.

If one projects on valence states of fixed $L^z$, the result is a one dimensional equation in the LF radial variable $\zeta$ where $\zeta^2 = b^2_\perp x(1-x)$ is conjugate to the $q \bar q$ LF kinetic energy $k^2_\perp\over x(1-x)$, which is also the square of the $q \bar q $ invariant mass 
${\cal M}^2_{q \bar q} = (p_q+p_{\bar q})^2$ for massless quarks: 
\begin{equation}
 [-{d^2 \over d\zeta^2} + {1-4L^2 \over \zeta^2} + U({\zeta^2}) ]\Psi_{q \bar q}(\zeta) = M^2_H \Psi_{q\bar q}(\zeta)
 \label{LFSE}
\end{equation}
where $L= {\it max} |L^z|$. The confining potential  $U(\zeta)^2 $ in this effective two-body equation reflects not only the usual QCD interactions, but also the effects from the elimination of the higher Fock states.
However, following the procedure of dAFF, the LF potential can only have form $U(\zeta)^2 = \kappa^4 \zeta^2 + {\it const} $, if one demands that the action of the semiclassical theory remain conformally invariant.   This procedure not only introduces the hadronic mass scale $\kappa$  but it  also ensures that only confined colored states can emerge.    The masses of each hadron consisting of light quarks and features such as  decay constants and the inverse of the 
slope of hadronic form factors will  also scale with $\kappa.$  
It should be emphasized that the absolute value of $\kappa$  in units such as MeV is not determined by chiral QCD -- only 
dimensionless ratios are predicted by the theory.  Thus in a sense, the model has no parameters.

\begin{figure}[h]
\centering
\includegraphics[width=0.80\textwidth]{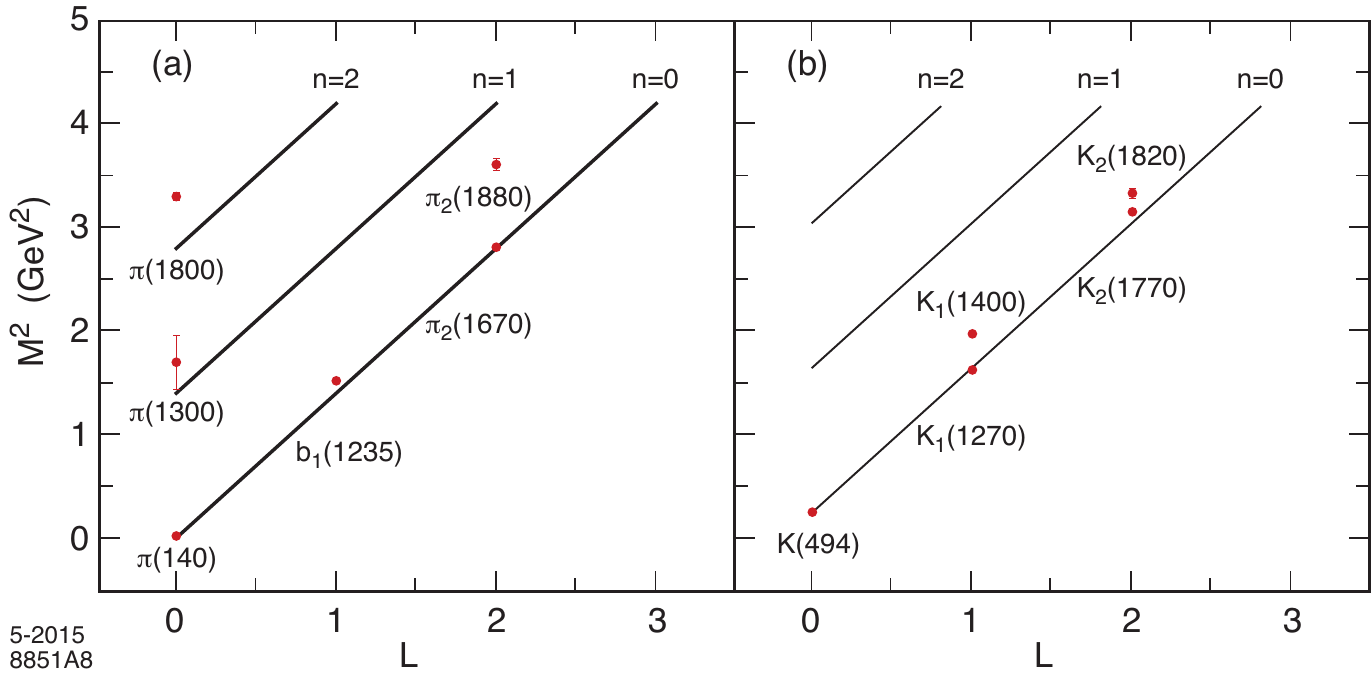}
\includegraphics[width=0.80\textwidth]{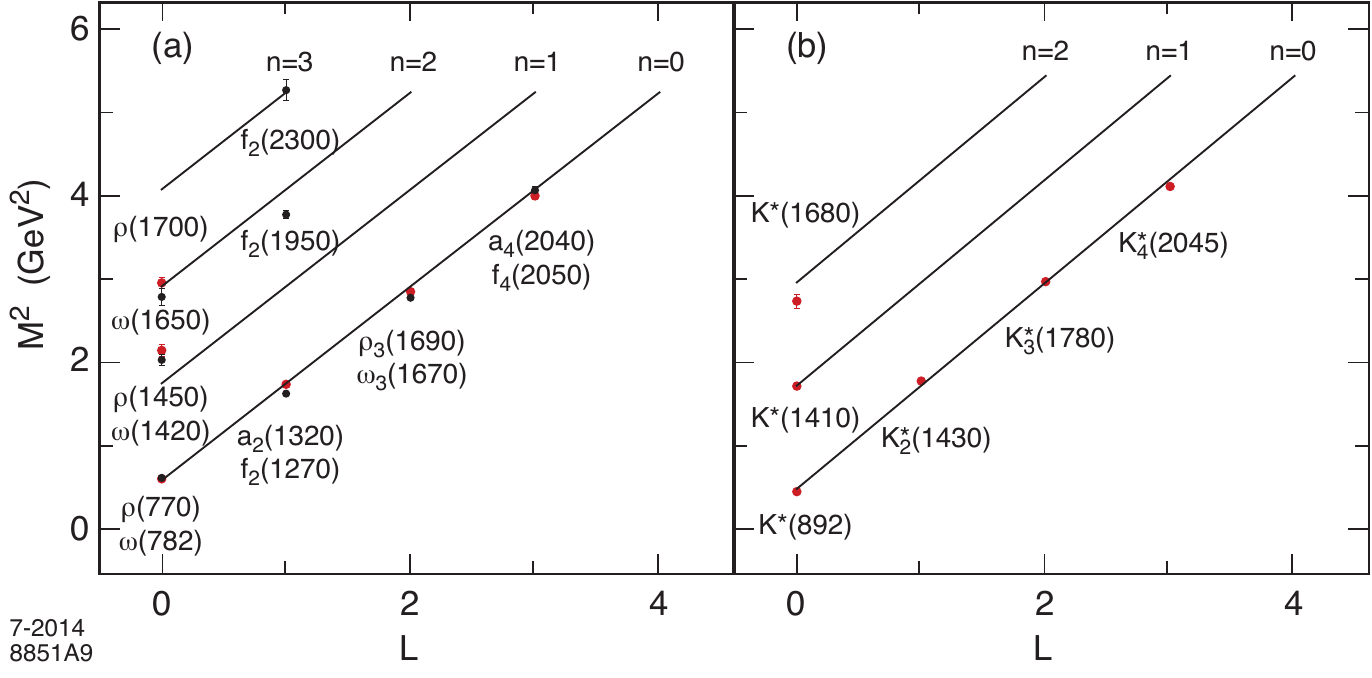}
\caption{\label{Fig:SlovakiaFig2}  
The predicted $S=0$  and $S=1$  meson trajectories in $n$ and $L$ from LF holography.}
\end{figure}

Remarkably, the identical bound state equations appear in the formalism built on AdS$_5$~\cite{Brodsky:2014yha} -- the five-dimensional representation of the isometries of the conformal group.   
In fact, because of {\it light-front holography }~\cite{Brodsky:2003px}, the equations derived from the AdS$_5$ action are identical to LF equations in $3+1$ space-time  at fixed LF time $\tau$ when one identifies the fifth dimension  coordinate $z$ of AdS$_{5}$  space with the LF transverse coordinate $\zeta$.  The Polchinski-Strassler formulae for electromagnetic and gravitational form factors  in  AdS$_{5}$  coincide with the Drell-Yan West LF formulae, an important feature of light-front holography~\cite{Brodsky:2007hb}.  
The identical potential emerges:  $U(\zeta)^2 = \kappa^4 \zeta^2 $ together with a   $J$-dependent constant $2\kappa^2(J-1)$  for spin-$J$ meson 
representations~\cite{deTeramond:2013it} if one modifies the AdS$_{5}$ action by  the ``soft-wall" dilaton factor $e^{+\kappa^2 z^2}.$  Thus the combination of the soft wall AdS/QCD model, light-front holography, and the dAFF procedure for retaining the conformal invariance of the action in LF Hamiltonian theory leads to the LF Schr\"odinger equation with the emerging scale $\kappa$ illustrated Fig. \ref{SlovakiaReduction}.   The eigenvalues of this LF Hamiltonian give a simple form for the meson spectrum for massless quarks: $M^2(n,L,J) = 4\kappa^2 (n+ {L+ J\over 2}) = 4\kappa^2 (n+ L+ {S\over 2}). $  
A comparison of the predicted spectrum with the observed $J=0$ and $J=1$  meson trajectories consisting of  $u,d,s $ quarks is shown in Fig. \ref{Fig:SlovakiaFig2}.
The predictions for hadrons containing strange quarks are obtained by including the quark mass term $\sum_i {m^2_i\over x_i}$ in the LF kinetic energy. 
%
%

\section{Features of the LF Schr\"odinger Equation}

The LF Schr\"odinger Equation has a number of remarkable features for both hadronic spectroscopy and dynamics.  

\begin{figure}[h]
\centering
\includegraphics[width=0.70\textwidth]{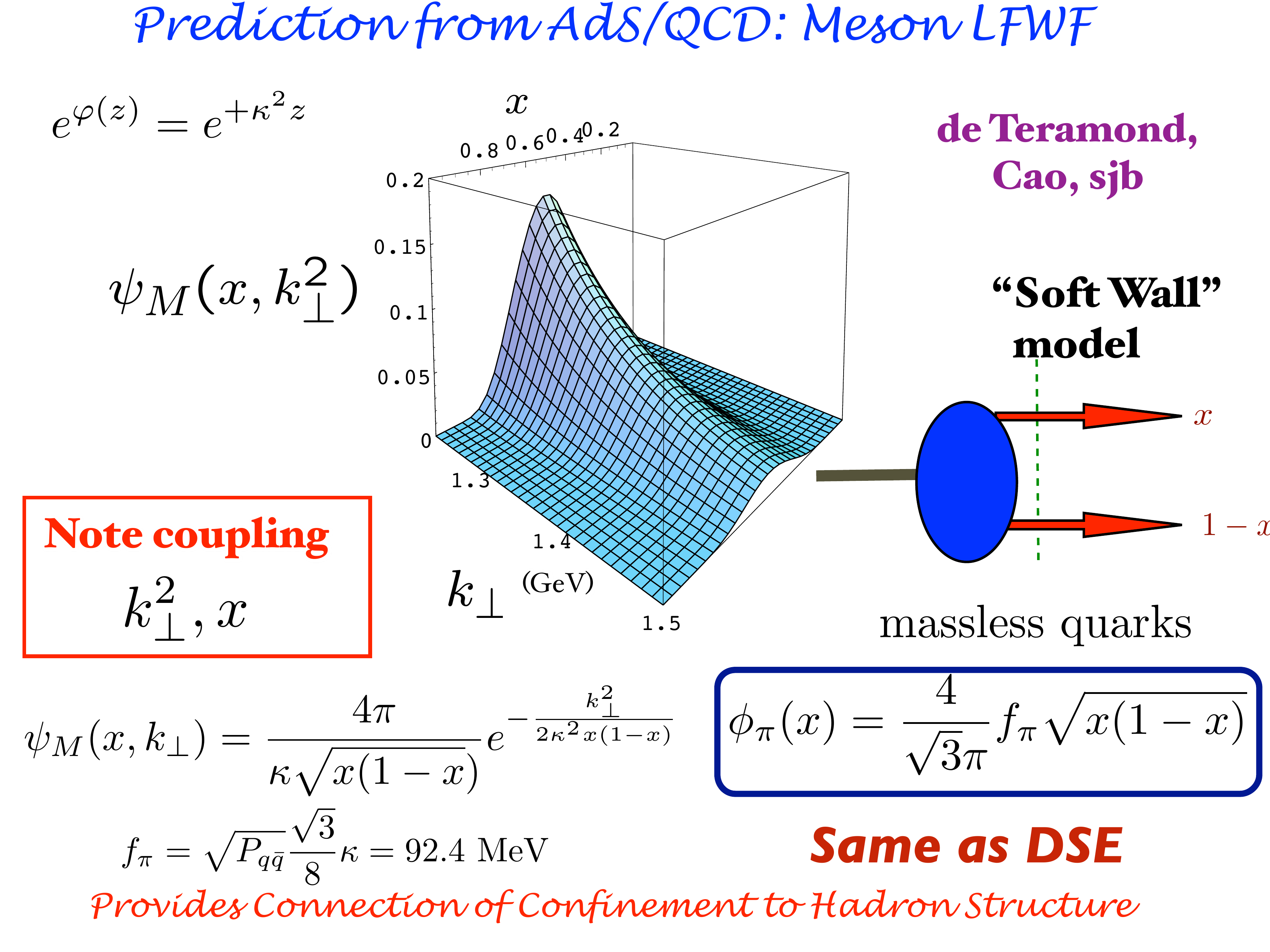}
\caption{\label{Fig:SlovakiaFig7}  
The  LF wavefunctions  and distribution amplitudes of mesons predicted by the  LF Schr\"odinger Equation.}
\end{figure}

\begin{itemize}

\item The pion eigenstate $(n=L=J =0)$   is massless for zero quark mass.   The positive terms from the LF kinetic energy and the confining LF harmonic oscillator potential are precisely cancelled by the negative constant term.  The physical pion mass corresponds to $m_q= 54$ MeV.

\item The predicted Regge trajectories are linear in $n$ and $L$ with the identical slope: ${\cal M}^2 = 4\kappa^2 ( n+ L + S/2) $

\item The  ${\cal M}^2$ of the $S=1$  vector meson  and  $ S=0$  pseudoscalar trajectories are separated  by $2\kappa^2$  for all $n$ and $L$.

\item The eigenfunctions determine the hadronic LF wavefunctions $\psi(x,k_\perp)$ in the nonperturbative domain which in turn determine hadron form factors, structure functions, etc. The transverse momentum  dependence $k_\perp$ always appears in the rotationally invariant combination $k^2_\perp/ x(1-x)$~\cite{Brodsky:1982nx}. 
See Fig. \ref{Fig:SlovakiaFig7}.

\item
The predicted  LF wavefunction  of the $\rho$ meson provides an excellent description of the empirical features of $\rho$ 
electroproduction as shown by Forshaw and Sandapen~\cite{Forshaw:2012im}.

\item The predicted form for the pion distribution amplitude $\phi_\pi(x) = {4 f_\pi \over \sqrt 3 } \sqrt{x(1-x) }$ in the nonperturbative domain leads to a prediction for the photon-to-pion transition form factor in agreement with Belle data~\cite{Brodsky:2011xx}. The $\sqrt{x(1-x)}$  dependence agrees with Dyson-Schwinger determinations~\cite{Chang:2014lva}.  It evolves to the $x(1-x)$ asymptotic form~\cite{Lepage:1979zb,Lepage:1980fj} as $\log Q^2 \to \infty$ using ERBL evolution.

\item Although the potential between light quarks is a harmonic oscillator, the corresponding nonrelativistic  potential $V(r) $ between heavy quarks is linear in $r$, consistent with quarkonium phenomenology. This connection~\cite{Trawinski:2014msa}  is due to the fact that the eigenvalue in ${\cal M}^2$ is effectively the square of the energy eigenvalue in the $\vec P  =0$ rest frame.

\end{itemize}

\section{Superconformal Algebra and Supersymmetric Features of QCD}

\begin{figure}[h]
\centering
\includegraphics[width=0.7\textwidth]{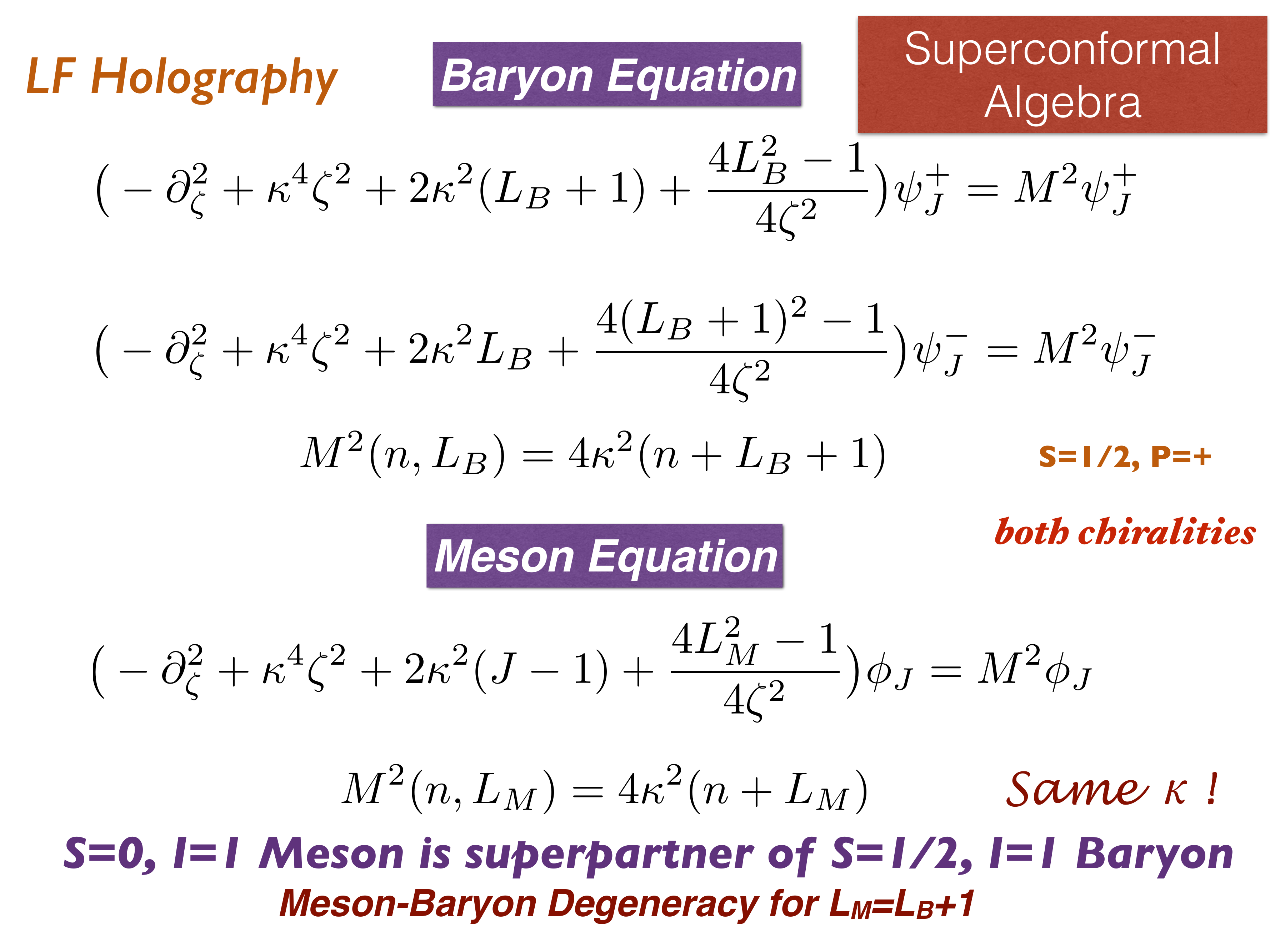}
\caption{\label{Fig:SlovakiaFigSuperconformal}  
The effective LF bound state equations for baryons and mesons with $m_q=0$ derived from superconformal algebra.   
The LF  wavefunctions $\psi_J^{\pm}$ refer to the quark spin $S^z_q$  parallel or anti-parallel to the parent baryon spin $S^z_B$.   The two amplitudes have the same mass squared eigenvalues and equal probabilities.}
\end{figure}

The dAFF procedure can be generalized to baryon states~\cite{deTeramond:2014asa} for massless quarks using the $2 \times 2$ Pauli matrix representation of the conformal group~\cite{Fubini:1984hf}. 
The potential in the LF equations are generated by the superconformal generator $S$ where the special conformal operator $K = \{S,S^\dagger\}$.
The resulting  two-component LF equations describe baryons as a quark plus a scalar diquark with the quark  $S^z_q$ parallel or anti-parallel to the parent baryon's $S^z_B,$
 as well as the LFSE for mesons. See Fig. \ref{Fig:SlovakiaFigSuperconformal}.

\begin{figure}[h]
\centering
\includegraphics[width=0.7\textwidth]{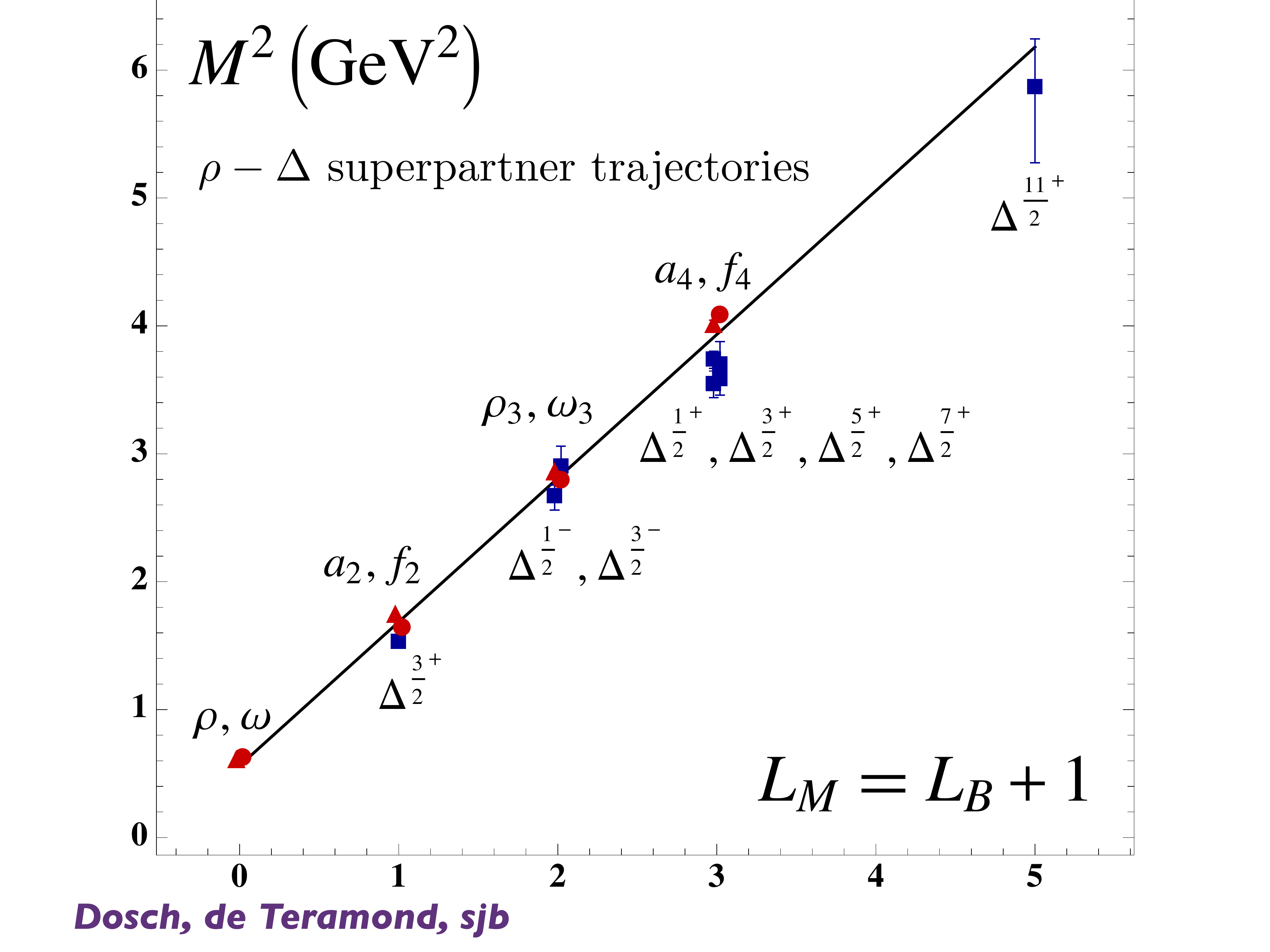}
\caption{\label{SlovakiaFig5}  A comparison of meson and baryon Regge trajectories.}
\end{figure}

The eigensolutions have the remarkable feature that  the entire baryon and meson spectra are degenerate if $L_M = L_B+1$;  i.e. the mesons  and baryons are 
superpartners if one makes  a shift of the internal angular momentum of the mesons by one unit of orbital angular momentum.  
See Fig. \ref{SlovakiaFig5}.
The pion and other mesons with $L=0$ have no superpartners.  
The empirical baryon and 
meson trajectories have the same slope in $n$ and $L$,
as predicted by holographic
QCD.  In
fact, the best fits to the numerical values for the Regge slopes
agree within $\pm10\% $ for all hadrons, mesons and baryons; this
leads to a near-degeneracy of  the meson and baryon levels in the
model.

\begin{figure}[h]
\centering
\includegraphics[width=0.9\textwidth]{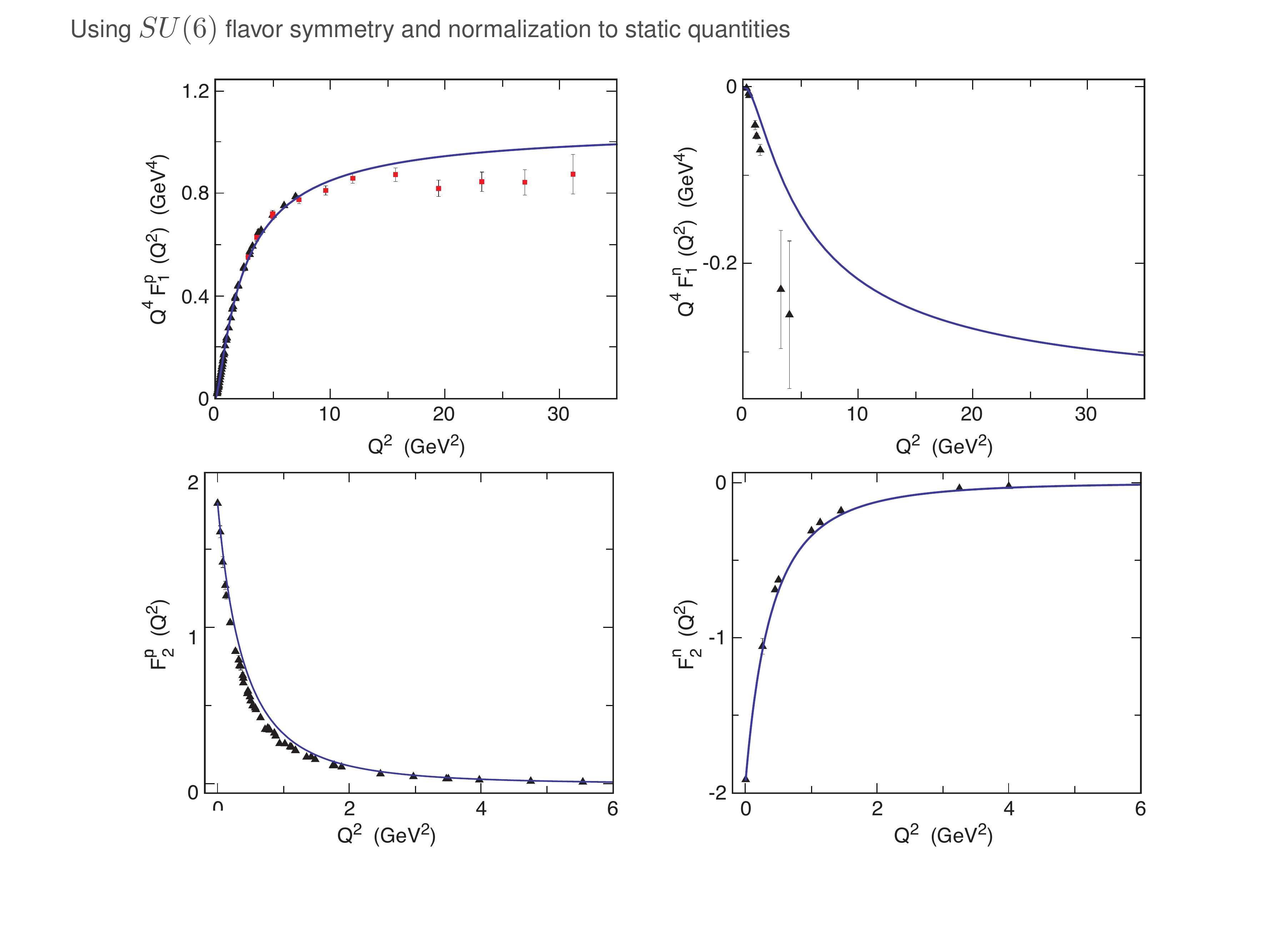}
\caption{\label{SlovakiaFig9}  Predictions for the nucleon Dirac and Pauli form factors. The Pauli form factors are normalized to the nucleon anomalous magnetic moments.}
\end{figure}

The nucleon form factors predicted by this formalism are shown in Fig. \ref{SlovakiaFig9}. 
Since the LF wavefunctions of mesons and baryons with $L_M = L_B+1$ and have same mass and  twist, one expects that their respective form factors will also be identical.

A generalization to heavy quark states is given in Ref.~\cite{Dosch:2015bca}. The idea to apply supersymmetry to hadron physics is not new~\cite{Miyazawa:1966mfa,Catto:1984wi,Lichtenberg:1999sc}.
 In~\cite{Miyazawa:1966mfa} mesons and baryons are grouped together in a large supermultiplet, a representation of  $U_{6/21}$. 
 In Ref. ~\cite{Catto:1984wi}
 the supersymmetry results of Miyazawa~\cite{Miyazawa:1966mfa}  are recovered in a QCD framework, provided that a diquark configuration emerges through an effective string interaction. 

\section{The Non-Perturbative QCD Coupling}

As noted above, the color confining potential of the LF Hamiltonian can be derived from the modification of the AdS space 
curvature,  the soft-wall dilaton profile~\cite{Brodsky:2013ar} $e^{+\kappa^2 z^2}$ which is  determined 
by the form of the potential dictated by the dAFF mechanism.
This same profile  leads to an explicit form for  the QCD coupling at long distances using  LF holography~\cite{Brodsky:2010ur,Deur:2015aka,Brodsky:2014jia}.
We will apply the AdS/QCD prediction to $\alpha_{g_1}^{AdS}(Q^2)$,   the effective charge in the $g_1$-scheme~\cite{Grunberg:1980ja}
defined from the Bjorken sum rule~\cite{Bjorken:1966jh,Bjorken:1969mm}. It is 
normalized to $\pi$ at $Q^2=0$ due to kinematical 
constraints~\cite{Deur:2005cf,Deur:2008rf}. This well-measured coupling can serve as the 
QCD-analog of the Gell-Mann-Low coupling $\alpha(Q^2)$ of QED~\cite{Brodsky:2010ur}.

The 5-dimensional AdS action is :
\begin{equation}
S=\frac{1}{4}\intop d^5 x  \sqrt{det(g^{AdS}_{\mu \nu})}~ e^{\kappa^{2}z^{2}} \frac{1}{g_{AdS}^{2}} F^{2}.
\end{equation}
The initially constant AdS coupling $\alpha_{AdS}\equiv g_{AdS}^{2}/4\pi$ can be redefined to
absorb the effects of the AdS deformation which determines the long-range confining interaction. It  predicts the form:
$g_{AdS}^{2}\rightarrow g_{AdS}^{2}~e^{\kappa^{2}z^{2}}$.   This constraint also  controls the form of $\alpha_s$ at small $Q^2$.  
Transforming to momentum space yields~\cite{Brodsky:2010ur}
\begin{equation} \label{alphaAdS}
\alpha^{AdS}_{g_1}(Q^2) = \pi  e^{- Q^2/ 4 \kappa^2}  .
\end{equation}

In pQCD, $\alpha_{s}\equiv {g_s^{2}/ 4\pi}$ acquires its large $Q^2$-dependence from short-distance
quantum effects. It is given by 
the QCD renormalization group equation, where the logarithmic derivative of the coupling defines the $\beta$ function.   
If $\alpha_s$ is small, one can use the perturbative expansion:
\begin{eqnarray} \label{beta}
Q^2{d \alpha_{s}}/{dQ^2} =
-(\beta_0 \alpha_s^2 + \beta_1 \alpha_s^3+ \beta_2 \alpha_s^4 +  \cdots).
\end{eqnarray}
The $\beta_i$ for $i  \ge 2$ are scheme-dependent  and are known up to order $\beta_3$  in the $\overline{MS}$ renormalization scheme.
Eq. (\ref{beta}) thus yields $\alpha_{\overline{MS}}(Q^2)$ at high $Q^2$.
In addition, 
$\alpha_{g_1}^{pQCD}(Q^2)$ can be expressed as a perturbative expansion in 
$\alpha_{\overline{MS}}(Q^2)$~\cite{Bjorken:1966jh,Bjorken:1969mm}.
Thus, pQCD predicts the form of  $\alpha_{g_1}(Q^2)$ with asymptotic freedom at large $Q^2$.

\begin{figure}[h]
\centering
\includegraphics[width=0.7\textwidth]{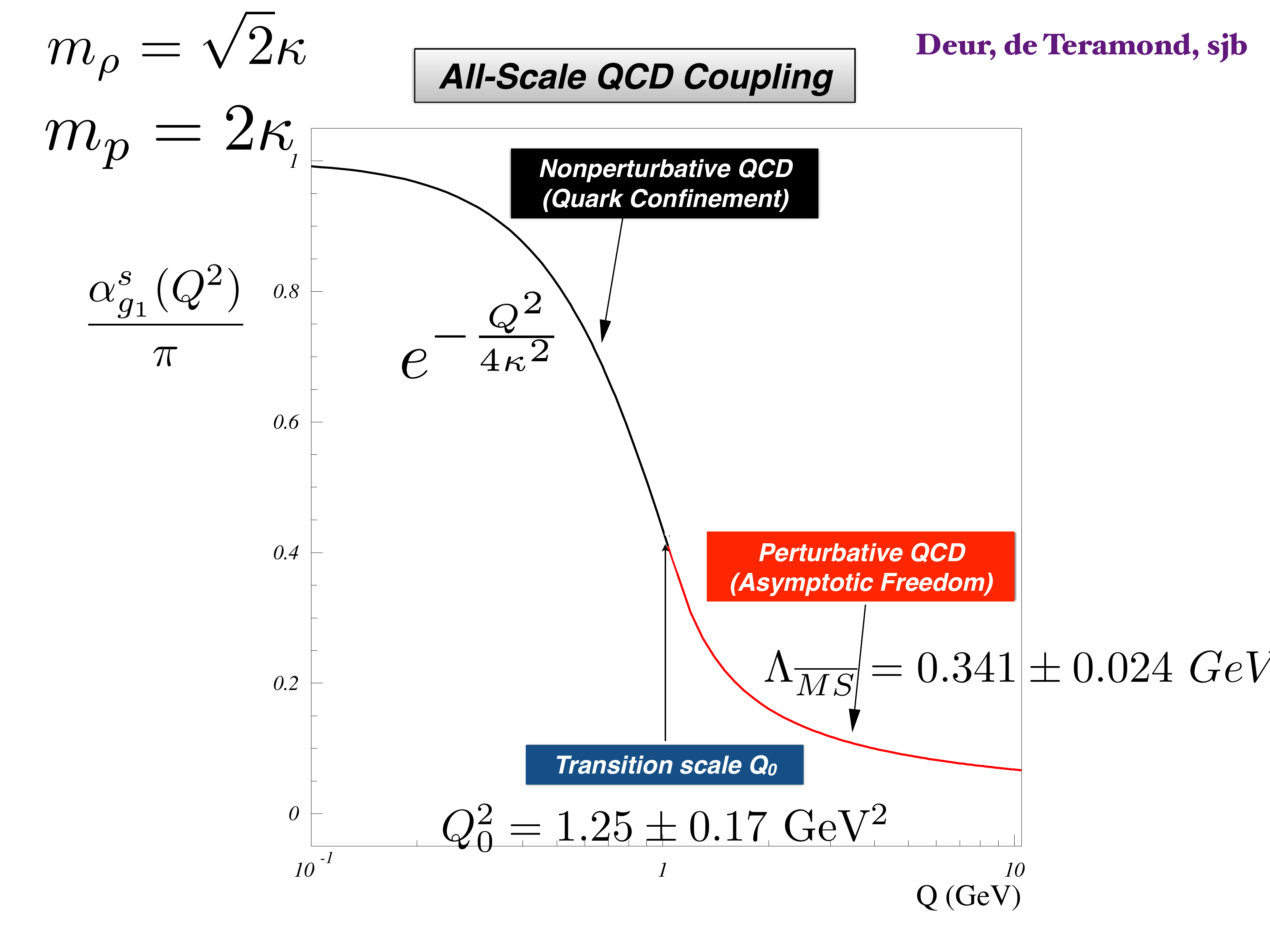}
\caption{\label{Matching}  
The strong coupling $\alpha_{g1}(Q^2)$ obtained from the analytic matching of perturbative and non-perturbative QCD regimes.}
\end{figure}

The existence at moderate values of $Q^2$ of a dual description 
of QCD in terms of either quarks and gluons versus  hadrons (``quark-hadron 
duality"~\cite{Bloom:1970xb}) is consistent with the matching of $\alpha_{g_1}^{pQCD}$ to $\alpha_{g_1}^{AdS}$ at intermediate values of $Q^2$.

This matching can be done by  imposing continuity of both $\alpha_{g_1}(Q^2)$ and its first derivative, 
as shown in Fig. \ref{Matching}. The unique solution for the resulting two equalities determines the scale parameter of the running coupling
$\Lambda_{\overline{MS}}$ from  $\kappa$, and fixes the scale $Q_0$ 
characterizing the transition between the large and short-distance regimes of QCD.    The result at order $\beta_3$, the same order to which the experimental 
value of $\Lambda_{\overline{MS}}$ is extracted, is $\Lambda_{\overline{MS}}=0.341 \pm 0.024$ GeV. 
The uncertainty stems from the extraction of $\kappa$  from the $\rho$ or  proton masses and from a small contribution
from  ignoring the quark masses. 
This theory uncertainty is less or comparable to that of the experimental determinations, which combine to~\cite{Deur:2015aka}
$\Lambda_{\overline{MS}} = 0.339 \pm 0.016$ GeV.

We thus can determine $\Lambda_{\overline{MS}}$ from hadron masses: $\Lambda_{\overline{MS}} = 0.440~m_\rho  = 0.622 ~ \kappa$. 
The scale $Q_0^2 =1.25~GeV^2$ determines the transition between the non-perturbative and perturbative domains.  
It can serve as the starting point for DGLAP and ERBL evolution in the $ {\overline{MS}}$ scheme.

\begin{figure}[h]
\centering
\includegraphics[width=0.7\textwidth]{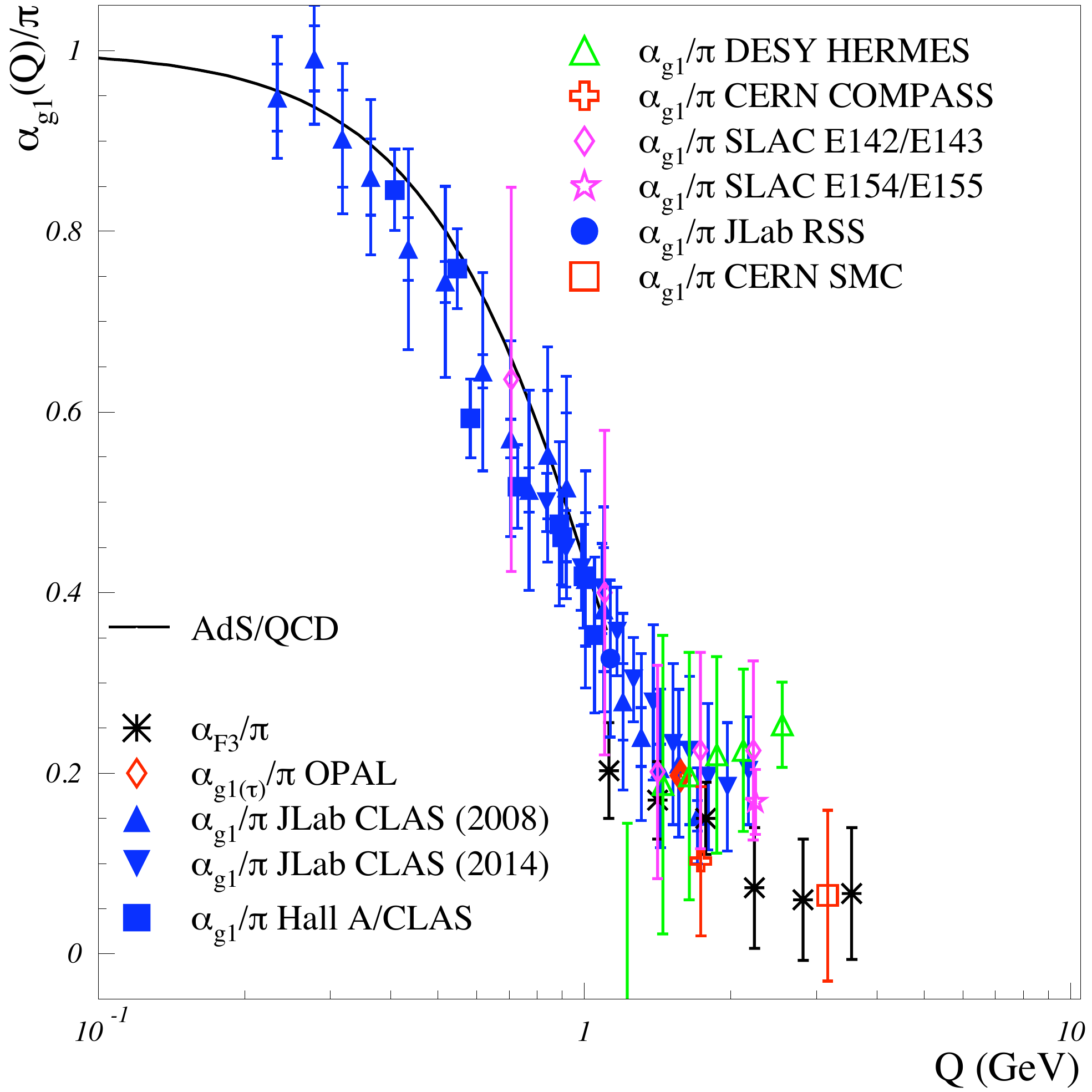}
\caption{\label{Fig:order dependence}  
The  analytic non-perturbative AdS/QCD prediction $\alpha^{AdS}_{g_1}(Q^2) = \pi  e^{- Q^2/ 4 \kappa^2} $ compared with measurements and the Bjorken sum rule.
 The data are plotted in the domain between $0<Q<Q_0$ where AdS/QCD is a valid approximation. }
\end{figure}

In  Fig. \ref{Fig:order dependence}, the AdS/QCD prediction for $\alpha^{AdS}_{g_1}(Q^2)$  
(\ref {alphaAdS}) is plotted together with data~\cite{Deur:2005cf,Deur:2008rf}. Even though 
there are no adjustable parameters, the predicted Gaussian form for the behavior of $\alpha^{AdS}_{g_1}(Q^2)$ at $Q^2 < Q^2_0$ agrees well with 
data~\cite{Brodsky:2010ur}.  There is little dependence of  
$\alpha^{pQCD}_{g_{1}}(Q^2)$ on the order of perturbation theory in $\beta_{n}$ and  $\alpha^n_{\overline{MS}}$.

\section{Hadronization at the Amplitude Level and other New Directions}

\begin{itemize}

\item
The new insights into color confinement given by AdS/QCD suggest that one could compute hadronization at  amplitude level~\cite{Brodsky:2009dr} using LF time-ordered perturbation theory, but including the confinement interaction.  For example, if one computes $e^+ e^- \to q \bar q \to q \bar q g \cdots$, the quarks and gluons only appear in intermediate states, and only hadrons can be produced.  LF perturbation theory 
provides a remarkably efficient method for the calculation of multi-gluon amplitudes~\cite{Cruz-Santiago:2015nxa}. 

\item
The eigensolutions of the AdS/QCD LF Hamiltonian can used to form an ortho-normal basis for diagonalizing the complete QCD LF Hamiltonian.  This method, ``basis light-front quantization"~\cite{Vary:2009gt}  is expected to be more efficient than the DLCQ method~\cite{Pauli:1985pv} for obtaining QCD 3+1  solutions.

\item
All of the hadron physics predictions discussed in this report are independent of the value of $\kappa$; only dimensionless ratios are predicted, such as $m_p = \sqrt 2 m_\rho$ and the ratio $\Lambda_{\overline  MS}/m_\rho$.    The ratio can be obtained in any renormalization scheme. 
One thus retains dilatation invariance   $\kappa \to \gamma  \kappa$ of the prediction..

\item
The $\kappa^4 \zeta^2$ confinement interaction between a $q$ and $\bar q$ will induce a $\kappa^4/s^2$ correction to $R_{e^+ e^-}$, replacing the $1/ s^2$ signal usually attributed to a vacuum gluon condensate.  

\item
The kinematic condition that all $k^+ = k^0+ k^3$ are positive and conserved precludes QCD condensate contributions to the $P^+=0$ LF vacuum state, which by definition is the causal, frame-independent lowest invariant mass eigenstate of the LF Hamiltonian~\cite{Brodsky:2009zd,Brodsky:2012ku}. 

\item
It is interesting to note that the contribution of the {\it `H'} diagram to $Q \bar Q $ scattering is IR divergent as the transverse separation between the $Q$  
and the $\bar Q$ increases~\cite{Smirnov:2009fh}.  This is a signal that pQCD is inconsistent without color confinement.  The sum of such diagrams could sum to the confinement potential $\kappa^4 \zeta^2 $ dictated by the dAFF principle that the action remains conformally invariant despite the mass scale in the Hamiltonian.

\end{itemize}

\section*{Acknowledgments}

Presented by SJB at {\it Hadron Structure Ô15},  Horny Smokovec, Slovak Republic, June 26 to July 3, 2015. 
We also thank Cecil Lorce and Kelly Yu-Ju Chiu,or helpful conversations.
This material is based upon work supported by the U.S. Department of Energy, Office of Science, Office of Nuclear Physics under contract DE--AC05--06OR23177. This work is also supported by the Department of Energy  contract DE--AC02--76SF00515.

\end{document}